\documentclass[aps,prl,superscriptaddress,twocolumn,letterpaper]{revtex4}
\usepackage{amsmath}
\usepackage{dcolumn}
\usepackage{epsfig}
\usepackage{graphicx}
\usepackage{latexsym}
\usepackage{amssymb}
\usepackage{color}
\usepackage{amsfonts}
\usepackage{bm}

\newcommand{\be}{\begin{eqnarray}}
\newcommand{\ee}{\end{eqnarray}}
\newcommand{\ba}{\begin{array}}
\newcommand{\ea}{\end{array}}

\begin{document}
\title{Particle-hole symmetry reveals failed superconductivity in the metallic phase of two-dimensional superconducting films}

\author{Nicholas P. Breznay}
\affiliation{Department of Applied Physics, Stanford University, Stanford, CA 94305, USA}
\affiliation{Department of Physics, Harvey Mudd College, Claremont, CA 91711, USA}
\author{Aharon Kapitulnik}
\altaffiliation{Corresponding author. Email: aharonk@stanford.edu}
\affiliation{Department of Applied Physics, Stanford University, Stanford, CA 94305, USA} 
\affiliation{Department of Physics, Stanford University, Stanford, CA 94305, USA}

\date{\today}

\begin{abstract}
Electrons confined to two dimensions display an unexpected diversity of behaviors as they are cooled to absolute zero. Noninteracting electrons are predicted to eventually ``localize'' into an insulating ground state, and it has long been supposed that electron correlations stabilize only one other phase: superconductivity. However, many two-dimensional (2D) superconducting materials have shown surprising evidence for metallic behavior, where the electrical resistivity saturates in the zero-temperature limit; the nature of this unexpected metallic state remains under intense scrutiny. We report electrical transport properties for two disordered 2D superconductors, indium oxide and tantalum nitride, and observe a magnetic field-tuned transition from a true superconductor to a metallic phase with saturated resistivity. This metallic phase is characterized by a vanishing Hall resistivity, suggesting that it retains particle-hole symmetry from the disrupted superconducting state.
\end{abstract}

\maketitle

\section{Introduction}
Conventionally, possible ground states of a disordered two-dimensional (2D) electron system at zero temperature include superconducting, quantum Hall liquid, or insulating phases. However, transport studies near the magnetic field-tuned superconductor-insulator transition in strongly disordered films suggested the emergence of anomalous metallic phases that persist in the zero-temperature limit, with resistances ($\rho_{xx}$) much lower than their respective nonsuperconducting state values $\rho_{N}$ (extrapolated from above the superconducting transition temperature $T_c$). Initial studies of such phases on amorphous MoGe films~\cite{Yazdani1995,Ephron1996,Mason1999} were followed by similar observations in amorphous indium oxide~\cite{Steiner2005,Liu2013}, tantalum~\cite{qin2006}, and indium-gold alloy~\cite{rosario2007} films, as well as in crystalline materials~\cite{tsen2016, saito2015} and hybrid systems consisting of a superconducting metal in contact with a two-dimensional electron gas such as tin-graphene~\cite{Han2014}. Metallic phases were also observed in weakly disordered 2D superconductors at zero field when disorder or carrier density are tuned~\cite{Han2014, Eley2012, couedo2016, Mason2001}.

Despite the ubiquitous appearance of this metallic phase, progress in understanding its origin has been slow. Early theoretical treatments explored quantum fluctuations in the presence of a dissipative bath, presumably due to residual fermionic excitations \cite{Shimshoni1998,Kapitulnik2001,Spivak2001,Feigelman2001,Goswami2006}, a Bose metal phase \cite{Das1999,phillips2003}, and an exotic non Fermi-liquid vortex metal phase \cite{Galitski2005,Mulligan2016}. Finally, noting that the distribution of the superconducting order parameter is highly inhomogeneous in the presence of disorder, Spivak {\it et al.} \cite{Spivak2008} examined a metallic phase that is stabilized by quantum fluctuations while showing significant superconducting correlations. To date there has been no conclusive evidence that distinguishes any one of these scenarios.

Here, we present evidence that the anomalous metallic phase can be described as a ``failed superconductor,'' where particle-hole symmetry, reminiscent of the superconducting state, plays a major role in determining its properties. This conclusion is a result of extensive Hall effect measurements on amorphous tantalum nitride (TaN$_x$) and indium oxide (InO$_x$) films that are weakly disordered~\cite{Steiner2008}. Specifically, we find that $\rho_{xx}$ in both systems becomes finite at the transition from a ``true superconductor'' to an anomalous metal at a magnetic field $H_{\textrm{M1}}$. The Hall resistance $\rho_{xy}$, zero in the superconductor because of electron-hole symmetry, remains zero for a wide range of magnetic fields, before becoming finite at a field $H_{\textrm{M2}}$ well below $H_{\textrm{c2}}$, the superconducting critical field. This apparent electron-hole symmetric behavior may herald the appearance of what has been termed the ``elusive'' Bose Metal~\cite{Das1999,phillips2003}.

\begin{figure*}[htb!]
\centering
\includegraphics[width=1.5\columnwidth]{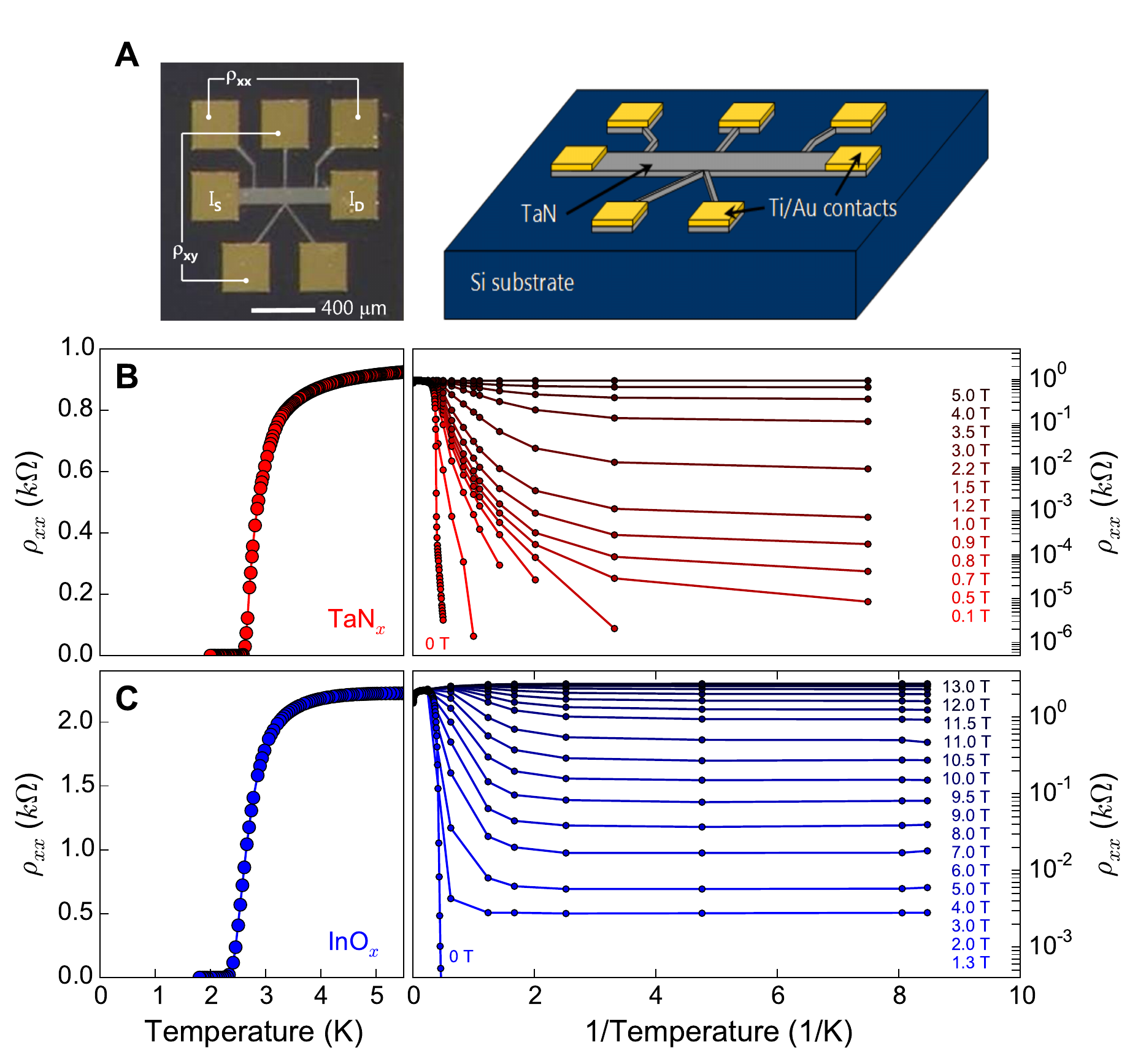}
\caption{Electrical transport in disordered superconducting devices. (A) Micrograph and schematic diagram of InO$_x$ and TaN$_x$ Hall bar devices. (B and C) Resistive transitions for TaN$_x$ (B) and InO$_x$ (C) films. Left: Zero-field resistivity versus temperature. Right: Resistive transitions in the indicated magnetic field plotted against inverse temperature, highlighting the saturated regime.}
\label{saturation}
\end{figure*}

\section{Results}
Figure~\ref{saturation} depicts a set of resistive transitions at increasing magnetic fields measured on amorphous TaN$_x$ and InO$_x$ films (pictured in Fig.~\ref{saturation}A). Sample growth and characterization details are described in the Supplementary Materials and in previous studies~\cite{Breznay2012,Breznay2013}. For TaN$_x$ the transition to saturated resistance evolves smoothly, such that for magnetic fields above $\sim$1\,T, saturation of the resistance is apparent; the lower field transitions seem to continue at lower temperatures in an activated fashion similar to MoGe~\cite{Ephron1996}. However, for InO$_x$, the transition from an activated behavior with a true superconducting state to a state with saturation of the resistance is more dramatic. Here, the resistance of the sample becomes immeasurably small below $\sim 1$\,K for magnetic fields below 1.2\,T ($T_c \approx 2.6$\,K for this sample.) In both materials the saturation persists to high fields and resistances comparable to the normal state resistance. However, Hall effect measurements indicate a sharp boundary at $H_{\textrm{M2}}$ between the anomalous metallic phases with $\rho_{xx} \ll \rho_{N}$ and the metallic behaivor that persists at higher fields.

\begin{figure*}[htb!]
\centering
\includegraphics[width=1.5\columnwidth]{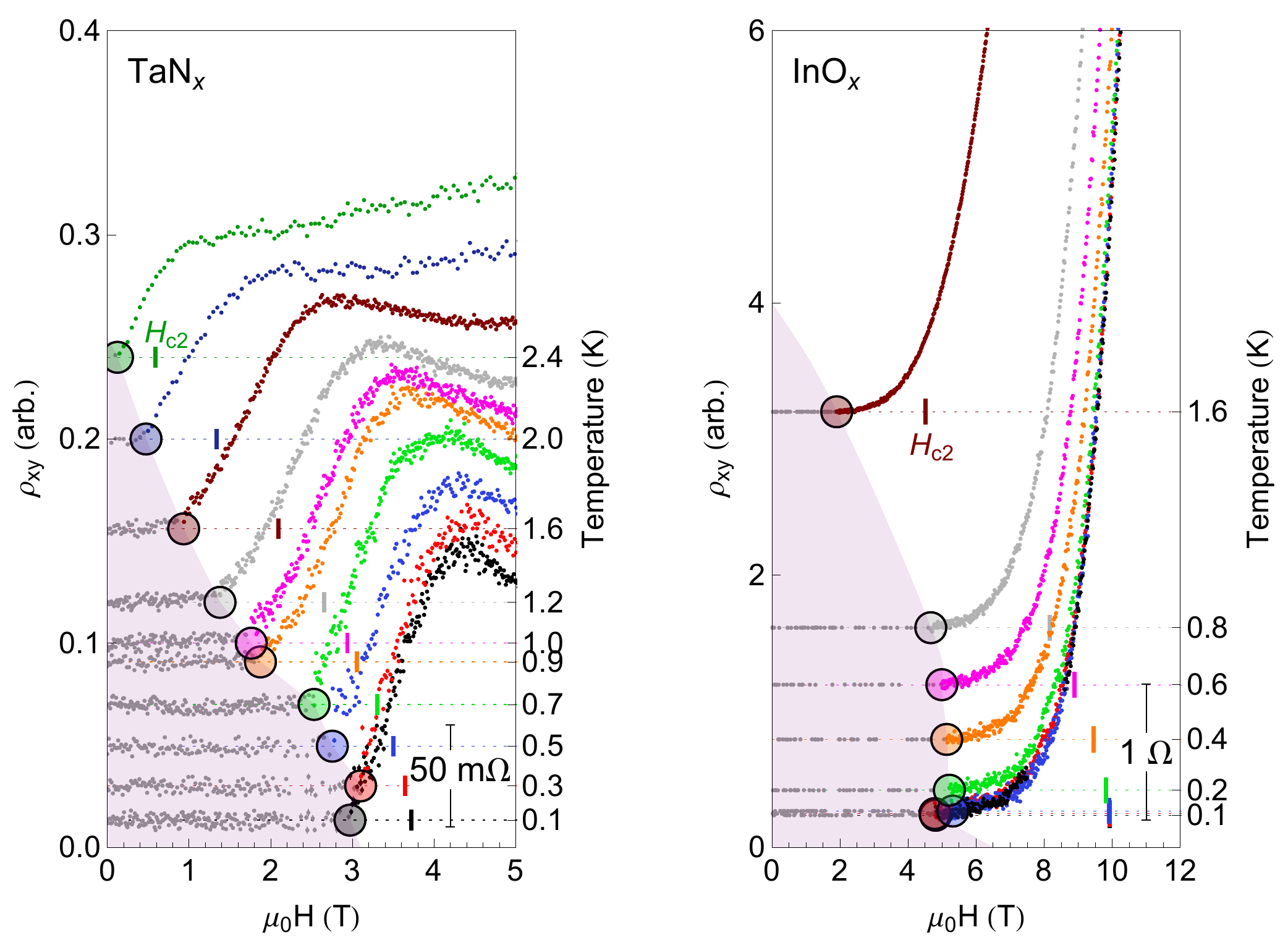}
\caption{Region of zero Hall effect. Hall resistivity versus temperature for weakly disordered TaN$_x$ (left) and InO$_x$ (right) films. The curves are offset vertically according to their temperature; the shaded region indicates where $\rho_{xy} = 0$ as a function of temperature and magnetic field, and an approximate location of $H_{\textrm{c2}}$ is marked for each curve. Scale bars for $\rho_{xy}$ are shown at the lower right.}
\label{hall}
\end{figure*}

Insight into the exotic nature of the anomalous metallic phase is obtained when we examine the behavior of the Hall effect at low temperatures, depicted for both, TaN$_x$ and InO$_x$ films in Fig.~\ref{hall}. Whereas in strongly disordered materials, the Hall resistance was found to be zero below the superconductor-insulator transition crossing point at $H_c$ (realized by InO$_x$ films~\cite{Paalanen1992,Breznay2016}), the weakly disordered films here show $\rho_{xy}=0$ up to a field $H_{\textrm{M2}} < H_{\textrm{c2}}$ (circled in Fig.~\ref{hall}). Furthermore, the Hall resistance is found to be zero (to our noise limit $\delta\rho_{xy}$, below which we cannot rule out a finite but very small $\rho_{xy}$) in a wide range of magnetic fields, $H_{\textrm{M1}} < H < H_{\textrm{M2}}$, where saturation of the longitudinal resistance is also observed. For TaN$_x$ the uppper limit $\delta\rho_{xy}$ is $\sim 3 \times 10^{-4}\ \Omega$, whereas for InO$_x$, the upper limit is $\delta\rho_{xy} \sim  5\times 10^{-4}\ \Omega$.

\begin{figure*}[htb!] 
\centering
\includegraphics[width=1.5\columnwidth]{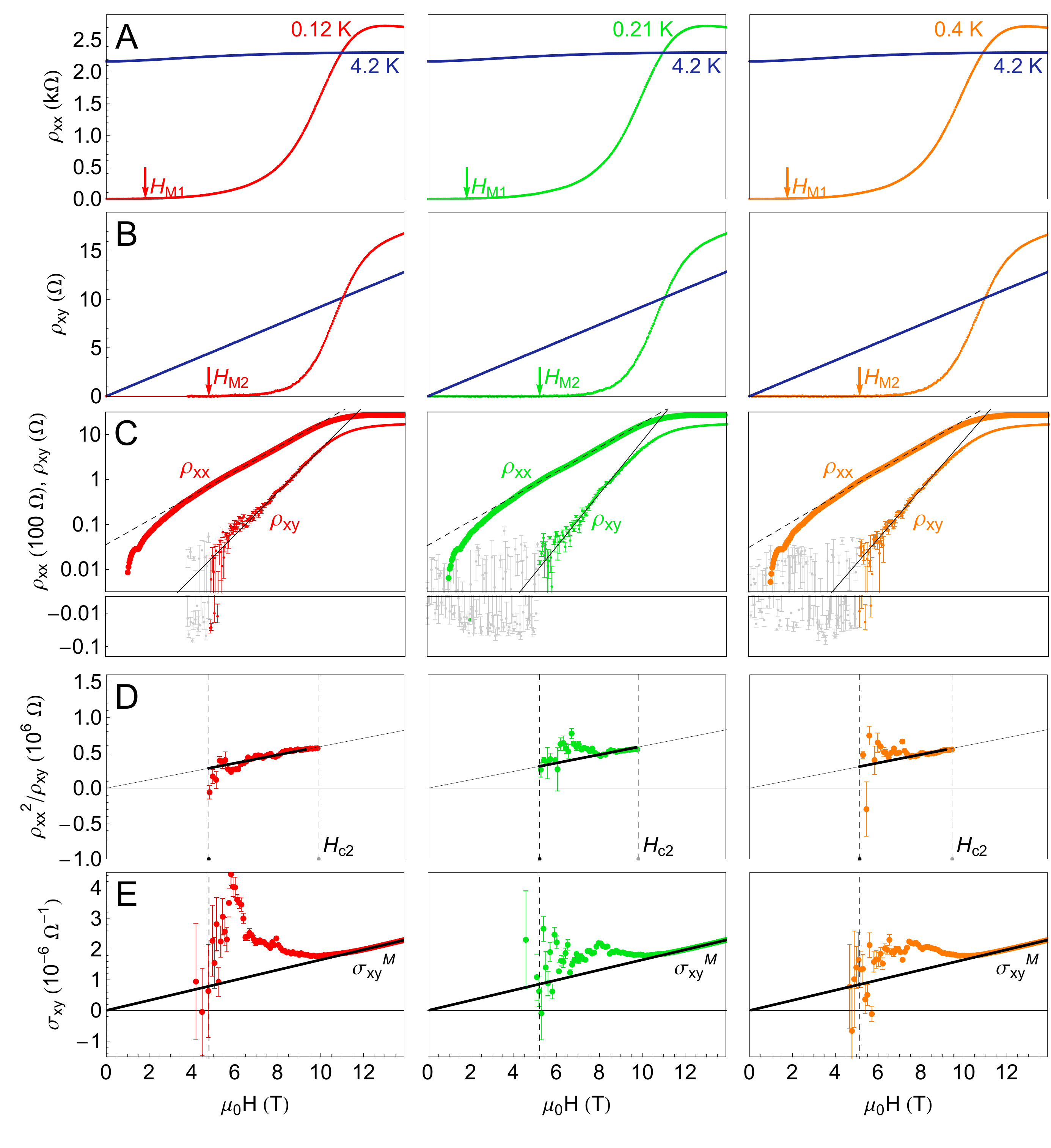}
\caption{Transport regimes in magnetic field. Longitudinal resistance (A), transverse resistance (B), scaling of $\rho_{xx}$ and $\rho_{xy}$ (C), the ratio $\rho_{xx}^2/\rho_{xy}$ (D), and $\sigma_{xy}$ (E) plotted versus magnetic field for three different temperatures for the InO$_x$ sample shown in Fig.~\ref{hall}. (A and B) Normal-state (4.2\,K) curves, for reference. (C) Regions of linear scaling (lines are guides to the eye). The transition from a true superconductor to an anomalous metallic phase is marked at $H_{\textrm{M1}}$, the transition from the anomalous metallic phase to a vortex flow-dominated superconductor at $H_{\textrm{M2}}$, and the mean field transition to the normal state at $H_{\textrm{c2}}$. The solid lines in (C) show the dissipative region of vortex motion (see main text); the solid line in (E) shows the normal-state metallic Hall conductivity $\sigma_{xy}^M$.}
\label{vinokur}
\end{figure*}

To further elucidate the fact that indeed there is a phase transition (or a sharp crossover at zero temperature) in the vortex state that appears at a field $H_{\textrm{M1}}$, we examine the nature of the vortex resistivity tensor in the entire field range below $H_{\textrm{c2}}$ in Fig.~\ref{vinokur}. Vortex motion should obey the scaling relation $\rho_{xy}\propto \left( \rho_{xx}^2/H \right) \tan\theta_H$~\cite{Vinokur1993}, whether exhibiting flux flow, thermally assisted flux flow, or vortex glass (creep) behaviors. 

Figure~\ref{vinokur} shows the longitudinal resistance $\rho_{xx}(H)$, the Hall resistance $\rho_{xy}(H)$, $\rho_{xx}$ and $\rho_{xy}$ with a log scale, the ratio $\rho_{xx}^2/\rho_{xy}$, and the Hall conductivity $\sigma_{xy}$ as a function of the magnetic field for InO$_x$ at temperatures below 1\,K. (Data for TaN$_x$ are presented in the Supplementary Materials.) These curves capture all three field-tuned transitions in the films. First, the longitudinal resistance $\rho_{xx}$ shows the transition to the metallic state at $H_{\textrm{M1}}$ (Fig.~\ref{vinokur}A), above which the Hall resistance $\rho_{xy}$ is still zero (Fig.~\ref{vinokur}B). Above $H_{\textrm{M2}}$, the Hall resistance is finite, and both $\rho_{xx}$ and $\rho_{xy}$ show $\rho \sim \exp(H/H_0)$ scaling (Fig.~\ref{vinokur}C), previously associated with the metallic phase in MoGe. In addition, above $H_{\textrm{M2}}$, $\rho_{xx}$ and $\rho_{xy}$ obey scaling of $\rho_{xx}^2/\rho_{xy} \propto H$ (Fig.~\ref{vinokur}D), indicating a state of dissipating vortex motion~\cite{Vinokur1993}. As a result, the Hall conductivity $\sigma_{xy}$ (Fig.~\ref{vinokur}E) starts to decrease with decreasing field and extrapolates to zero at $H_{\textrm{M2}}$, further supporting the picture of a particle-hole symmetric state. Because $\rho_{xx}$ extrapolates to a finite value at zero temperature in fields above $H_{\textrm{M2}}$, we identify this regime with a pure flux flow resistance. Finally, as we increase the field beyond $H_{\textrm{c2}}$ both $\rho_{xx}$ and $\rho_{xy}$ recover their normal-state values. 

The identification of zero $\rho_{xy}$ in the anomalous metallic regime needs to be tested against the possibility that it is just too low to measure because of the appearance of local superconducting ``puddles''. In particular, because the anomalous metal-superconductor system is expected to be inhomogeneous~\cite{Shimshoni1998,Spivak2008}, we may have a system of superconducting islands (for which $\sigma_{xx}^S \rightarrow \infty$ and $\sigma_{xy}^S =0$) embedded in a metal (characterized by $\sigma_{xx}^M$ and $\sigma_{xy}^M$). If the metal percolates, then for any dilution of the system by superconducting ``islands'' the measured Hall conductivity satisfies $\sigma_{xy}= \sigma_{xy}^M$~\cite{Stroud1984}; this behavior would persist until the superconductivity is quenched. In Fig.~\ref{vinokur}E, we plot $\sigma_{xy}^M$ (thick lines) along with $\sigma_{xy}$ calculated by inverting the resistivity tensor: $\sigma_{xy}=-\rho_{xy}/[\rho_{xx}^2+\rho_{xy}^2]$. Above $H_{\textrm{c2}}$, $\sigma_{xy}$ shows normal state behavior. However, just below $H_{\textrm{c2}}$ and well above $H_{\textrm{M2}}$, $\sigma_{xy}$ has departed from $\sigma_{xy}^M$, indicating that the anomalous metallic state (as well as the vortex liquid phase above it) is not simply a matrix of superconducting ``puddles'' embedded in a metal matrix.

\section{Discussion}
Before we discuss the resulting phase diagram for these 2D disordered films, several points need to be emphasized.  First, in the absence of superconducting attractive interactions, these films are expected to be weakly localized and insulating in the limit of zero temperature, although this limit is impossible to observe in finite-sized films with good metallic conduction. Second, the phases that we probe are all identified at finite magnetic fields and finite temperatures. In principle, in the presence of a finite magnetic field, there are no true finite-temperature superconducting phases in two dimensions in the presence of disorder~\cite{Giamarchi1995}, whereas in practice, the superconducting phase that we identify exhibits zero resistance. The transition to this phase, either as a function of temperature or magnetic field through $H_{\textrm{M1}}$, is therefore understood as a sharp crossover to a state with immeasurably low resistance. In a similar way, we understand the anomalous metallic phase that exhibits zero hall resistance. At low temperatures, this Hall resistance persists to be zero through $H_{\textrm{M1}}$ (presumably continuing to manifest electron-hole symmetry) but abruptly becomes finite above $H_{\textrm{M2}}$.

\begin{figure}[htb!] 
\centering
\includegraphics[width=1.0\columnwidth]{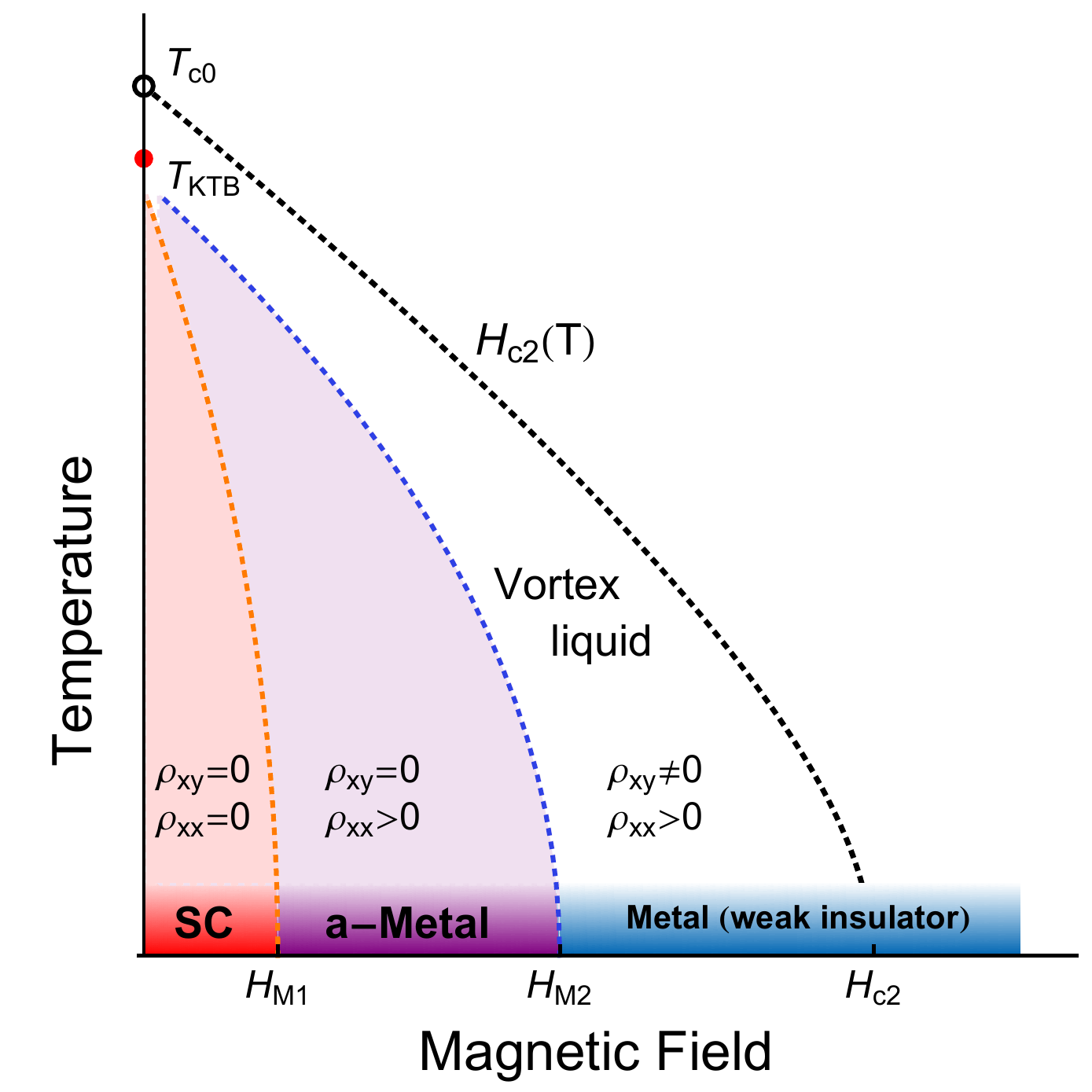}
\caption{Schematic phase diagram for a weakly disordered 2D superconductor. In zero field, a true superconducting (SC) state with transition temperature $T_{\textrm{KTB}}$ is manifested by zero resistance (see main text). Increasing the magnetic field uncovers a transition to an anomalous metallic (a-Metal) phase at $H_{\textrm{M1}}$, a transition to a vortex-flow dominated superconductor at $H_{\textrm{M2}}$, and the mean field transition to the normal state at $H_{\textrm{c2}}$ (and $T_{c0}$ in zero field). Dashed lines, extracted from observed transitions in the longitudinal and/or transverse resistances, represent finite-temperature crossovers. True phase boundaries lie at $H = 0$ and in the limit of zero temperature.}
\label{pd}
\end{figure}

In Fig.~\ref{pd} we show the resulting phase diagram for the two systems studied here. At low magnetic field, we observe a narrow superconducting phase, characterized by $\rho_{xx}$ that decreases exponentially with decreasing temperature and zero Hall effect. Upon increasing the magnetic field, the resistance of the sample starts to show saturation in the limit of $T \rightarrow 0$, while the Hall effect does not seem to change from zero. We identify this anomalous metallic phase (``a-Metal'' in the figure) already at a finite temperature, where saturation starts to be pronounced, but the important feature here is the extrapolation to zero temperature where true phase transitions manifest at a finite magnetic field. By increasing the magnetic field, we observe a vortex liquid phase, which is commonly observed in measurements on superconducting films at finite temperature. Although these results suggest the low temperature metallic phase proposed by Spivak {\it et al.}~\cite{Spivak2008}, this proposed phase is highly inhomogeneous and deserves further exploration. Assuming that a metallic phase needs a connection to a dissipative bath, an inhomogeneous state is a likely scenario~\cite{Shimshoni1998,Kapitulnik2001}.

The anomalous metallic phase identified above seems to exhibit strong superconducting pairing character, but no finite superfluid density on the macroscopic scale. This is seen in the experiments of Liu {\it et al.} \cite{Liu2013} where the ac response of 2D low-disorder amorphous InO$_x$ films, comparable to those discussed here, exhibited a superconducting response on short length and time scales in the absence of global superconductivity. Similar reasoning leads to the conclusion that vortices can be defined on short length and time scales, similar to the considerations that led to the Kosterlitz-Thouless transition~\cite{Kosterlitz1972}, where the superfluid density vanishes above the transition at $T_{\textrm{KTB}}$, but vortex-antivortex pairs are observed to proliferate through the system. 

The activated part of the resistive transition, just above saturation, fits a 2D collective vortex creep behavior. Hence, it is expected that by lowering the temperature toward $T=0$, saturation is a consequence of a change in vortex transport, such as a transition to a dissipation-dominated quantum tunneling~\cite{Mason1999,Kapitulnik2001}. Finally, by building on recent connections between more strongly disordered films and the quantum Hall liquid-to-insulator transition~\cite{Breznay2016}, we here observe that the metallic region can be described as an analog to the composite Fermi liquid observed in the vicinity of half-filled Landau levels of the 2D electron gas~\cite{Mulligan2016}.

\section{Materials and methods}
\subsection{Sample growth and characterization}
Disordered InO$_x$ films were grown using electron-beam deposition onto cleaned silicon substrates with silicon oxide; careful control of the sample growth resulted in amorphous, nongranular films (33). Films of TaN$_x$ were deposited using a commercial reactive sputtering tool (AJA International) onto plasma-etched silicon substrates. Film thicknesses (5~to~10\,nm) were confirmed by x-ray reflectivity and transmission electron microscopy. In both materials, the films can be considered 2D with respect to superconductivity and localization effects. Film compositions ($x \approx 1.5$ for InO$_x$ and $x \approx 1$ for TaN$_x$) were checked via x-ray reflectivity, diffraction, and photoemission spectroscopy; we adopted the notation of ``InO$_x$'' and ``TaN$_x$'' throughout the text as a reminder that the films are amorphous and nonstoichiometric. Film homogeneity was characterized using transmission electron microscopy and scanning electron microscopy, as well as optically; we found no evidence of inhomogeniety or granularity on any length scale to below the film thickness. Hall bar devices with a width of 100\,$\mu$m and aspect ratios of either 2 or 4 were fabricated using conventional photolithography techniques, with argon-ion milling to define the bar structure and electron-beam evaporated Ti/Au contacts with thicknesses of 10/100\,nm.

\subsection{Measurement and data analysis}
We measure the longitudinal resistance $\rho_{xx}$ and Hall resistance $\rho_{xy}$ using conventional four-point low-frequency ($\approx 10$\,Hz) lock-in techniques; reported values are in the linear response regime. Magnetoresistance and Hall measurements were performed at both positive and negative fields; the Hall resistance was extracted from the transverse voltage by extracting the component antisymmetric in the magnetic field. Measurements on $>$10 samples for both materials were checked in multiple cryostats; data at temperatures below 2\,K used a commercial top-loading dilution refrigerator with 14\,T superconducting magnet. Error bars on the raw data represent $\pm1\sigma$.

\section{}

\begin{acknowledgments}
\textbf{Acknowledgments:} We acknowledge illuminating discussions with Boris Spivak and Steven Kivelson. \textbf{Funding:} Initial work was supported by the National Science Foundation grant NSF-DMR-9508419. This work was supported by the Department of Energy (grant DE-AC02-76SF00515). \textbf{Author contributions:} NPB and AK conceived and performed the experiments and wrote the manuscript. \textbf{Competing interests:} The authors declare no competing interests. \textbf{Data availability:} All data needed to evaluate the conclusions in the paper are present in the paper and/or the Supplementary Materials. Additional data may be requested from the authors.
\end{acknowledgments}


\begin{references}

\bibitem{Yazdani1995} 
A. Yazdani and A. Kapitulnik, 
Phys. Rev. Lett. 74, 3037 (1995).

\bibitem{Ephron1996}
D. Ephron, A. Yazdani, A. Kapitulnik, and M.R. Beasley, 
Phys. Rev. Lett. 76, 1529 (1996).

\bibitem{Mason1999} 
N. Mason and A. Kapitulnik, 
Phys. Rev. Lett. 82, 5341 (1999).

\bibitem{Steiner2005}
M.A. Steiner and A. Kapitulnik, 
Physica C 422, 16 (2005).

\bibitem{Liu2013}
W. Liu, L.-D. Pan, J. Wen, M. Kim, G. Sambandamurthy, and N. P. Armitage, 
Phys. Rev. Lett. 111, 067003 (2013). 

\bibitem{qin2006}
Y. Qin, C. L. Vicente, and J. Yoon, 
Phys. Phys. Rev. B 73, 100505(R) (2006).

\bibitem{rosario2007}
M. M. Rosario, H. Wang, Yu. Zadorozhny, Y. Liu, 
J. Low. Temp. Phys. 147, 623 (2007).

\bibitem{tsen2016}
A.W. Tsen, B. Hunt, Y. D. Kim, Z. J. Yuan, S. Jia, R. J. Cava, J. Hone, P. Kim, C. R. Dean and A. N. Pasupathy,
Nature Physics 12, 208 (2016).

\bibitem{saito2015}
Yu Saito, Yuichi Kasahara, Jianting Ye, Yoshihiro Iwasa, and Tsutomu Nojima,
Science 350, 409 (2015).

\bibitem{Han2014}
Z. Han, A. Allain, H. Arjmandi-Tash, K. Tikhonov, M. Feigel'man, B. Sac\'ep\'e and V. Bouchiat, 
Nature Physics 10, 5 (2014).


 \bibitem{Eley2012}
S. Eley, S.G. Gopalakrishnan, P. M. Goldbart, N. Mason, 
Nature Physics 8, 59 (2012).

\bibitem{couedo2016}
F. Couëdo, O. Crauste, A. A. Drillien, V. Humbert, L. Bergé, C. A. Marrache-Kikuchi, and L. Dumoulin, 
Sci. Rep. 6, 35834 (2016).

\bibitem{Mason2001}
Nadya Mason and Aharon Kapitulnik, 
Phys. Rev. B 64, 060504(R) (2001).

\bibitem{Shimshoni1998}
E. Shimshoni, A. Auerbach and A. Kapitulnik, 
Phys. Rev. Lett. 80, 3352 (1998).

\bibitem{Kapitulnik2001}
A. Kapitulnik, N. Mason, S.A. Kivelson, and S. Chakravarty, 
Phys. Rev. B 63, 125322 (2001).

\bibitem{Spivak2001}
B. Spivak, A. Zyuzin, and M. Hruska, Phys. Rev. B 64, 132502 (2001).

\bibitem{Feigelman2001}
M.V. Feigel'man, A.I. Larkin and M.A. Skvortsov, Phys. Rev. Lett. 86, 18691872 (2001).

\bibitem{Goswami2006}
P. Goswami and S. Chakravarty, Phys. Rev. B 73, 094516 (2006).

\bibitem{Das1999}
D. Das and S. Doniach, Phys. Rev. B 60, 1261 (1999).

\bibitem{phillips2003}
Philip Phillips and Denis Dalidovich.
Science 302, 243 (2003).

\bibitem{Galitski2005}
V. M. Galitski, G. Refael, M. P. A. Fisher, and T. Senthil, Phys. Rev. Lett. 95, 077002 (2005).

\bibitem{Mulligan2016}
M. Mulligan, S. Raghu, Phys. Rev. B 93, 205116 (2016).

\bibitem{Spivak2008}
B. Spivak, P. Oreto, and S.A. Kivelson,  
Phys. Rev. B 77, 214523 (2008).

\bibitem{Steiner2008}
Myles A. Steiner, Nicholas P. Breznay, and Aharon Kapitulnik, 
Phys. Rev. B 77, 212501 (2008).

\bibitem{Breznay2012}
N. P. Breznay, K. Michaeli, K. S. Tikhonov, A. M. Finkel'stein, M. Tendulkar, and A. Kapitulnik,
Phys. Rev. B 86, 014514 (2012).

\bibitem{Breznay2013}
N. P. Breznay and A. Kapitulnik,
Phys. Rev. B 88, 104510 (2013).

\bibitem{Paalanen1992}
M. A. Paalanen, A.F. Hebard, and R.R. Ruel, 
Phys. Rev. Lett. 69, 1604 (1992).

\bibitem{Breznay2016}
Nicholas P. Breznay, Myles A. Steiner, Steven A. Kivelson, Aharon Kapitulnik,
PNAS 113, 280 (2016).

\bibitem{Vinokur1993}
V. M. Vinokur, V. B. Geshkenbein, M. V. Feigel'man, and G. Blatter, 
Phys. Rev. Lett. 71, 1242 (1993).

\bibitem{Stroud1984}
D. Stroud and D.J. Bergman, 
Phys. Rev. B 30, 447 (1984).

\bibitem{Giamarchi1995}
T. Giamarchi and P. Le Doussal,
Phys. Rev. B 52, 1242 (1995).

\bibitem{Kosterlitz1972}
J.M. Kosterlitz and D.J. Thouless,
Journal of Physics C: Solid State Phys. 5, L124 (1972).

\bibitem{Kowal}
D. Kowal, Z. Ovadyahu,
Solid State Commun. 90, 783 (1994).

\end{references}
\end{document}